# Multifunctional Metasurface Design with a Generative Adversarial Network

*Sensong An[1], Bowen Zheng[1], Hong Tang[1], Mikhail Y. Shalaginov[2], Li Zhou[1], Hang Li[1], Myungkoo Kang[3], Kathleen A. Richardson[3], Tian Gu[2], Juejun Hu[2], Clayton Fowler[1,\*], Hualiang Zhang[1,\*]*

[1]Department of Electrical & Computer Engineering, University of Massachusetts Lowell, Lowell, MA 01854, USA
[2]Department of Materials Science & Engineering, Massachusetts Institute of Technology, Cambridge, MA 02319, USA
[3]CREOL, University of Central Florida, Orlando, Florida 32816, USA
E-mail: clayton_fowler@uml.edu, hualiang_zhang@uml.edu

Abstract: Metasurfaces have enabled precise electromagnetic wave manipulation with strong potential to obtain unprecedented functionalities and multifunctional behavior in flat optical devices. These advantages in precision and functionality come at the cost of tremendous difficulty in finding individual meta-atom structures based on specific requirements (commonly formulated in terms of electromagnetic responses), which makes the design of multifunctional metasurfaces a key challenge in this field. In this paper, we present a Generative Adversarial Networks (GAN) that can tackle this problem and generate meta-atom/metasurface designs to meet multifunctional design goals. Unlike conventional trial-and-error or iterative optimization design methods, this new methodology produces on-demand free-form structures involving only a single design iteration. More importantly, the network structure and the robust training process are independent of the complexity of design objectives, making this approach ideal for multifunctional device design. Additionally, the ability of the network to generate distinct classes of structures with similar electromagnetic responses but different physical features could provide added latitude to accommodate other considerations such as fabrication constraints and tolerances. We demonstrate the network's ability to produce a variety of multifunctional metasurface designs by presenting a bifocal metalens, a polarization-multiplexed beam deflector, a polarization-multiplexed metalens and a polarization-independent metalens.

Metasurfaces, the two dimensional (2D) versions of metamaterials, are planar/conformal devices composed of subwavelength structures, called meta-atoms,[1, 2] which are capable of tailoring amplitude, phase, polarization and angular momentum of incident waves.[3-5] By manipulating the geometry of the individual meta-atom, independent phase and amplitude control of electromagnetic (EM) field can be achieved. Recently, metasurfaces consisting of all-dielectric meta-atoms have drawn enormous attention, [4, 6, 7] due to their unique capability of supporting EM multipole resonances and significantly lower losses as compared to their metallic counterparts.[8-11] The multipole responses in a meta-atom can be highly complicated, even for simple shapes, and thus an arbitrary meta-atom's response to an incident EM wave is difficult to predict. Traditional design approach relies on empirical reasoning or trial-and-error,[4, 7] which is inefficient and often ineffective, since this approach involves tremendous numerical full-wave simulations (e.g. finite-element method (FEM), finite-difference time-domain (FDTD) method and finite integration technique (FIT)), which provide accurate predictions but are extremely time consuming. Therefore, it can be time-consuming and laborious to find an appropriate set of meta-atoms for a specific design. Meanwhile, multifunctional metasurfaces such as multi-wavelength metasurfaces,[8, 9, 12, 13] multi-polarization metasurfaces[14] or reconfigurable metasurfaces based on phase change materials,[15-17] have presented another major design challenge due to the difficulty in exploring vast parameter spaces containing meta-atoms that possess sufficient complexity to satisfy restrictive design requirements. Therefore, in addition to exploring meta-atoms with basic shapes such as rings,[8, 9] cubes[13, 18] or cylinders,[19] previous works have also adopted evolutionary algorithms (EA) to search for meta-atoms and metasurface with free-form patterns[12, 14, 20, 21] which provides additional degrees of freedom (DOF). Nevertheless, the capability of this approach largely depends on the quality of the initial guess, which limited their stability and efficacy as the complexity of the problem grows.

To address the challenges in designing non-intuitive meta-atom and metasurfaces, several deep neural network (DNN) based approaches have been proposed and investigated since 2018. A "tandem" network structure[22-33] combining a pre-trained simulator with another model generator overcomes the issue of non-unique solutions (common to all inverse design problems), and allows the inverse neural network to converge steadily. However, there are several limitations to this approach. First, the designs constructed via tandem networks are dealing with variations of simple canonical geometries defined by a few parameters. The examples of such geometries include planar layers[22], metallic bars[25], dielectric spheres[26] and cylinders.[27] The lack of available DOF restricts functionality and performance of the designed metasurface. Training of the tandem networks can be extremely difficult, if not impossible, for meta-atoms with free-form geometries because of the dimension mismatch between the small number of EM response inputs (1D spectrum) and large design parameter outputs (2D images). Therefore, such tandem inverse design networks can only generate a single design based on each input setting, which suggests that the networks are "memorizing" the results rather than "learning" the design mechanisms and "composing" new solutions.

GANs provide a promising solution to mitigate these limitations. Since GANs were first introduced in 2014,[34] they have been widely applied in the field of image processing, due to their unique ability to reveal the hidden distributions behind enormous training datasets and compose complex and diverse designs based on limited inputs. Through Conditional Generative Adversarial Nets (CGAN)[35] (a GAN variant that is able to generate required designs conditioned on class labels), several design networks have been developed for realizing metallic metasurface filters with defined transmittance spectra[23] and dielectric meta-gratings with specified beam deflection angles and working frequencies.[36-38] However, these existing GAN-enabled metasurface design networks deal with either only amplitude responses [23, 39-43] or with reduced dimensionality structures (1D meta-grating designs at a supercell level [36, 37, 44]). Due to the hardly-predictable phase jumps caused by electromagnetic poles,[36] a GAN-enabled meta-atom design approach that deals with both amplitude and phase responses has been challenging and until this work, has remained an open question. This challenge significantly limit the efficacy of GAN-enabled design approaches, since most optical metasurfaces reshape the wavefront of the incident light by introducing local variations in both phase and amplitude.

In this work, we present a novel approach for designing free-form all-dielectric metasurface devices by combining the CGAN with the Wasserstein Generative Adversarial Networks (WGAN).[45, 46] As another widely-adopted GAN variant, WGAN introduces Earth-Mover distance as a loss evaluation method, which not only stabilizes the training process, but also qualifies the network for handling comprehensive metasurface design problems. Moreover, the proposed approach handles multiple inputs in parallel, meaning that the complexity is not affected by the size of the inputs. This feature further positions this GAN-based approach as the preferred solution in tackling multifunctional inverse design problems. Based on this highly efficient network, we have designed and verified several multifunctional metasurfaces, including a bifocal lens, a polarization-multiplexed beam deflector and two multifunctional metalenses in order to illustrate the versatility and scalability of the proposed method (some of these results are included in the Supporting Information). The presented examples substantiate that our approach demonstrates several important milestones as 1) the first free-form all-dielectric meta-atom design network; 2) the first free-form multifunctional metasurface design network and 3) the first metasurface lens designed by GANs.

**Network architecture.** The proposed network (**Figure 1**) combines the WGAN structure with the method of a CGAN, which trains a generator that maps a set of design conditions, $x$, combined with a Gaussian noise vector, $z$, to produce a target design (or fake samples), $y'$, defined as: $y' = G(z|x)$. We treated each meta-atom as a 2D image and all meta-atoms are of the same height. In general, the conditions $x$ can be any kind of auxiliary information. In this case, $x$ carries the electromagnetic responses obtained from numerical simulations of real samples $y$. The discriminator calculates the Wasserstein distance between the real samples $y$ and the generated fake ones $y'$, then it inversely tunes the parameters within the generator/ discriminator network to minimize/maximize the Wasserstein distance by using a gradient descent algorithm. The Wasserstein distance between a target design $y \in P_{data}$, and a generated one $y' \in P_G$, is defined as :

$$W(P_{data}, P_G) \approx \sup_{\|D\|_L \leq 1} \left\{ \mathbb{E}_{y \sim P_{data}}[D(y|x)] - \mathbb{E}_{y' \sim P_G}[D(y'|x)] \right\} \qquad (1)$$

where $P_{data}$ and $P_G$ are the sets of electromagnetic responses extracted from the training data and produced by the generator, respectively; $\mathbb{E}$ stands for the expected value and $D$ is the Wasserstein distance given by the discriminator.

The parameter optimizations for the generator and the discriminator are processed in turns, aiming for opposite objectives: the generator $G$ is trained to produce samples that cannot be distinguished by the discriminator, while the discriminator is trained to detect generated samples as fake. After the network is fully trained, the losses of both generator and discriminator, defined as:

$$L_g = -\mathbb{E}_{y' \sim P_G}[D(y'|x)] \qquad (2)$$

$$L_d = \mathbb{E}_{y' \sim P_G}[D(y'|x)] - \mathbb{E}_{y \sim P_{data}}[D(y|x)] \qquad (3)$$

are minimized and stabilized. Once both generator and discriminator are constructed with enough capacity, the stabilized losses indicate that their performance has plateaued, because $P_g$ already equals $P_{data}$. At this point, the generator can generate samples, $y'$, virtually identical to the real samples, $y$, (under the conditions $x$) such that the discriminator is unable to differentiate between them. Detailed flow diagrams of the discriminator and generator are shown in Fig. 1b. We used the Leaky ReLu-BatchNorm-transposed convolution modules[47, 48] for the generator, and the ReLu-BatchNorm-convolution modules for the discriminator (see Supporting Information Section 1 for detailed network architectures).

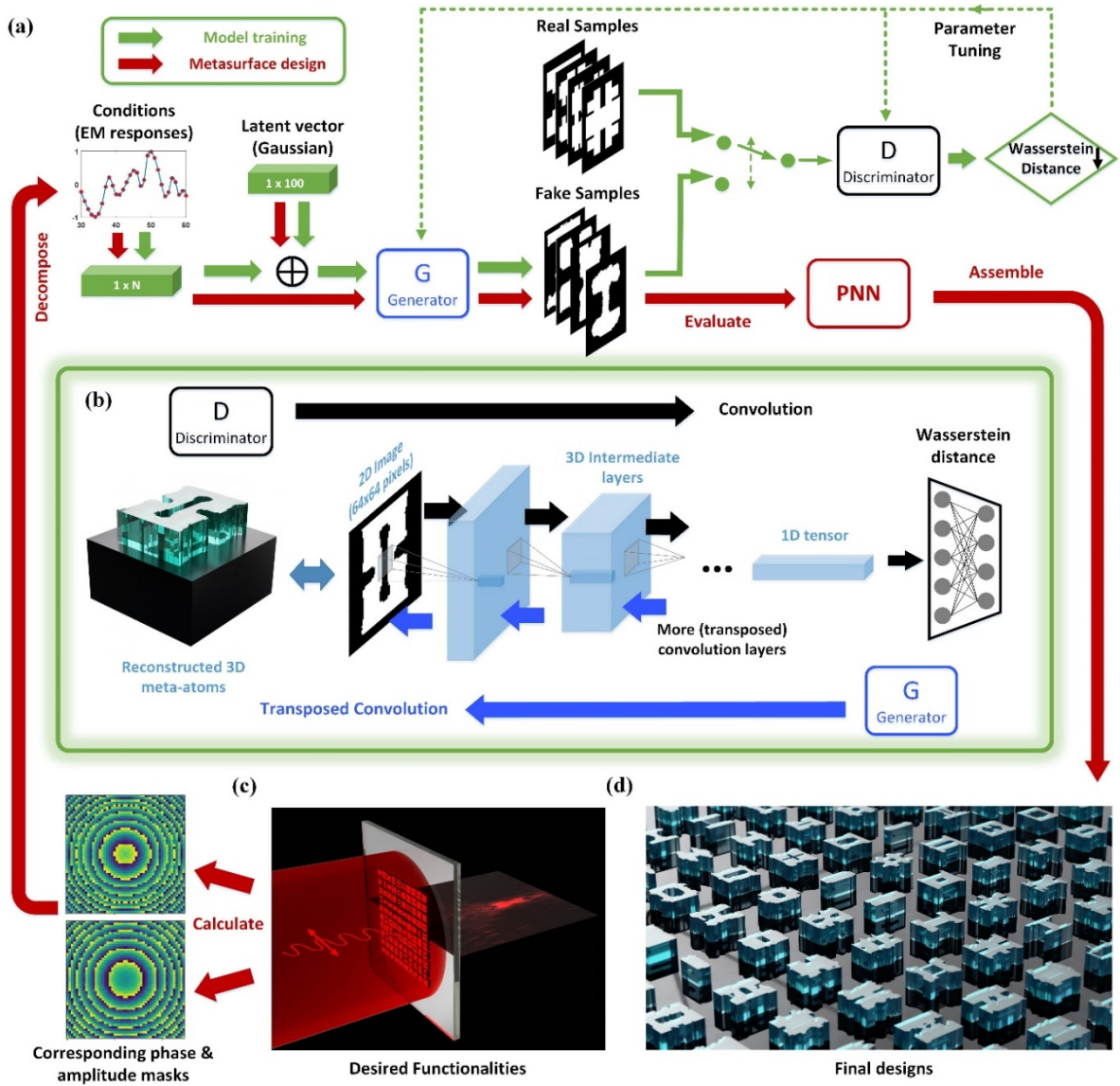

**Figure 1.** Network architecture of the generative meta-atom design network. (a) Schematic diagram of the proposed network. The discriminator network measures the Wasserstein distance between real and fake samples and aims to maximize the distance between them. The generator network attempts to confuse the discriminator by transforming target conditions combined with noise prior to producing fake samples that resemble real ones. Both components approach the ground truth through parameter tuning during this adversarial process. (b) Flow diagram of the generator and discriminator: detailed network structures, including kernel size and output tensor shapes are included in the Supporting Information, Section 1. (c) An example of the design process employing the well-trained model. The desired functionalities (lens, deflector, etc.) are translated into phase and amplitude masks, and later fed into the generator

to generate actual meta-atom arrays satisfying the design goals. (d) Rendered illustration of an exemplary metasurface design assembled with GAN-designed meta-atoms.

Without loss of generality, the all-dielectric meta-atom under consideration is made by patterning a 1-μm-thick film of dielectric material with a refractive index of 5 placed on a dielectric substrate with a refractive index of 1.4. The unit cell size was set to be $2.8 \times 2.8$ μm$^2$, which is designed to operate in the 5-10 μm spectral range (Fig. 1b, left side). Each meta-atom was generated with the "needle drop" approach to maximize the accessible pattern diversity (see Supporting Information Section 2). As part of the preprocessing, 2D cross-sections of all meta-atom from the training dataset were rescaled into $64 \times 64$ pixel images and binarized (dielectric parts to ones and voids to zeros) prior being fed into the discriminator for evaluation. The 2D dimensions of each image gradually decreased when passed through (2, 2) stride convolutional layers. The output of each layer is batch-normalized and passed through a ReLU activation function before it is sent to the next layer. The generator takes the conditioned prior noise $(z|x)$ as an input which is then provided to seven consecutive transposed convolutional layers. Each layer is followed by a leaky ReLU activation function for conditioned image generation. After the last transposed convolutional layer, a *tanh* activation function generates an image ready for evaluation. Finally, a pre-trained prediction neural network (PNN)[49] characterizes these output images and eliminates the unqualified meta-atom designs.

Conditions of the proposed generative network structure are formed with single or multiple correlated/uncorrelated EM targets, and therefore this network is uniquely applicable to multifunctional meta-atom/metasurface designs. To explore the potential of our network architecture, we constructed and trained a single wavelength meta-atom design network and a multifunctional metasurface design network, which were used for design and verification of a bifocal lens, a polarization-multiplexed deflector, a polarization-multiplexed lens and a polarization-independent lens (some of these devices are presented in the Supporting Information).

**Meta-atom design network.** A meta-atom design network was developed to generate meta-atom patterns based on conditioned phase and amplitude profiles following the proposed network architecture. Due to the phase-related design difficulties as mentioned in ref.[27], a preprocessing layer that translates relative phase and amplitude responses into complex transmission coefficients was added before the first layer of inputs. The real and imaginary parts of the transmission coefficients form the conditions $x$, such that $x = [T_{real}(y), T_{imag}(y)]$. Specifically customized for the meta-atom design task, a novel gradient penalty approach was also adopted to further stabilize the training process (see Supporting Information Section 3). Without loss of generality, the operating frequency was set to be 50 THz (6 μm wavelength). After 1,500 epochs of training, both the discriminator and generator losses were minimized and stabilized (Hyperparameters and training curves are included in Supporting Information Section 4), indicating that the network is fully trained. Unlike traditional GANs, which largely rely on the tuning of the hyperparameters to stabilize the training, the proposed network is highly stable and easily converges. Common training problems with GANs, like gradient explosions and vanishings, were not experienced during the training process regardless of the learning rate, optimizer type, number of layers, etc.

Several randomly selected phase and amplitude combinations were chosen to test the trained network model. For each phase and amplitude combination, we employed the well trained GAN to consecutively generate 100 qualified designs to check the generation stability and efficiency of the proposed approach. **Figure 2** presents several randomly selected phase and amplitude combinations (marked with red dots). The electromagnetic responses of the generated patterns were computed using full-wave simulation tools and are labeled with blue dots. We set a minimum threshold of ±0.1 amplitude error and ±10° phase error, as outlined by red lines in each polar plot of Fig. 2. The qualified designs for each target are highlighted in yellow, while designs with performance that fall outside the red outlines are left dark. As shown in Fig. 2, given the specific amplitude and phase targets, numerous qualified meta-atom designs of various shapes can be generated within a few seconds, manifesting the proposed design approach is highly time efficient. Having only one unqualified design among 600 extracted meta-atoms (Fig. 2c) indicates the network's high generalization stability and accuracy.

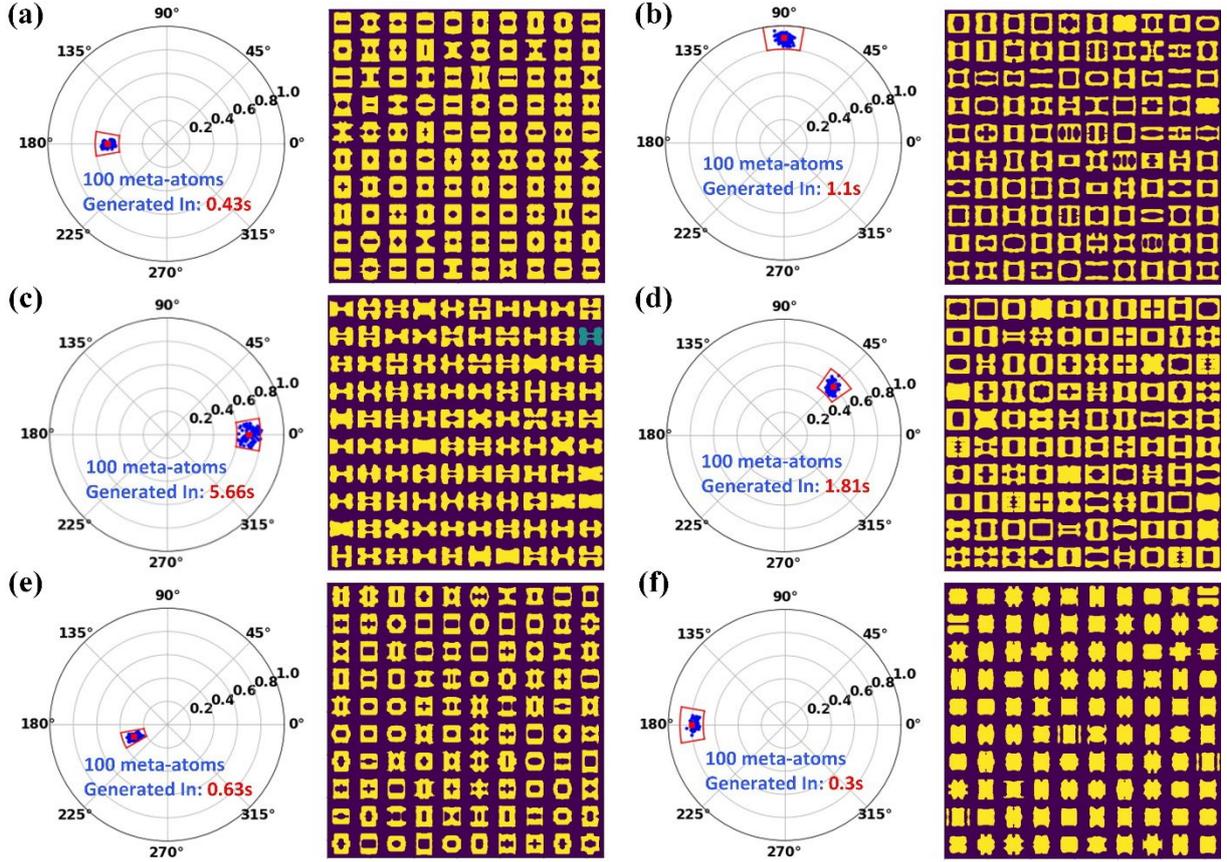

**Figure 2.** Meta-atom designs generated using a fully-trained conditional WGAN model. 100 meta-atom designs were produced for each combined condition of amplitude and phase: (a) 0.5 + 180°, (b) 0.9 + 90°, (c) 0.7 + 0°, (d) 0.6 + 45°, (e) 0.3 + 200° and (f) 0.8 + 180°, respectively. In the polar charts the radial and angular coordinates correspond to the transmission amplitude and phase. Blue dots represent EM responses of generated designs, red dots represent the targeted amplitude and phase conditions. 2D patterns of meta-atoms from each design group are shown on the right side of each subplot. Red outlines indicate the bounds of the allowed phase-amplitude values, while the corresponding qualified patterns are highlighted in yellow. Illumination of each meta-atom is performed from with an x-polarized plane wave from the substrate side.

The ability to design on demand meta-atom geometries with the specified phase and amplitude responses enables various striking applications, including multi-focal metalenses[9], beam deflectors[50], holograms[3] and airy beam generators[9, 11]. Using the trained network, we designed a bifocal metalens operating at 50THz, which requires simultaneous amplitude and phase modulations of each meta-atom within the metalens. To reduce the complexity of full-wave based verification process, the size of this metalens is limited to $50 \times 50$ meta-atoms (140 µm by 140 µm). For the designed bifocal lens with two focal points ($f_1$ and $f_2$) aligned laterally, the phase and amplitude profiles at $(x_0, y_0)$ on the metalens are given by:

$$a(x_0, y_0)e^{j\varphi(x_0, y_0)} = \frac{a_1}{d_1}exp^{j\frac{2\pi}{\lambda}d_1} + \frac{a_2}{d_2}exp^{j\frac{2\pi}{\lambda}d_2} \qquad (4)$$

where $d_1$ and $d_2$ are the distance between the point $(x_0, y_0)$ and two focal points $f_1$ and $f_2$, respectively, $a_1$ and $a_2$ are the amplitudes for the two focal points. In this demonstration, we set $f_1 = f_2 = 60$ µm and $a_1 = a_2 = 1$. Two focal points are separated by a lateral distance (along the x-axis) $d = 60$ µm. The calculated amplitude and phase masks are plotted in **Figure 3**a and 3b, respectively, illustrating the requirements for independent amplitude and phase control. In order to demonstrate the ideal (theoretical) performance of this dual focal lens, we theoretically calculated and plotted the electric field distribution in the X-Z plane at y = 0 (Fig. 3c) by modeling each meta-atom as a linearly polarized point source with the amplitude and phase values taken from Fig. 3a and 3b. The two $E_x$ field maxima occur as expected by the design at the two focal spots that are 60 µm away from the metasurface plane.

The amplitude and phase design targets for each meta-atom are designated as inputs to the GAN, while the output qualified meta-atom designs are used to assemble the whole device. The final metasurface design, along with the amplitude and phase profiles for each meta-atom as verified by numerical simulations, is shown in Fig. 3d and 3e, respectively. The electric field distribution produced by this metalens was also modeled using full-wave simulations (Fig. 3f). This design demonstration highlights the three key advantages of our approach: 1) over 600 qualified free-form meta-atoms were generated in less than 1 minute, indicating its time efficiency; 2) the excellent agreement between targeted (Fig. 3a, b) and simulated (Fig. 3e) meta-atom performances, as well as the agreement between the theoretical calculation (Fig. 3c) and the full-wave simulation result (Fig. 3f), validates its accuracy; and 3) considering almost all target functionalities can be decomposed into specific phase and amplitude requirements on the meta-atom level, this design approach can be easily extended for the design of various other devices including beam deflectors, holograms, etc., demonstrating that it is scalable and universal.

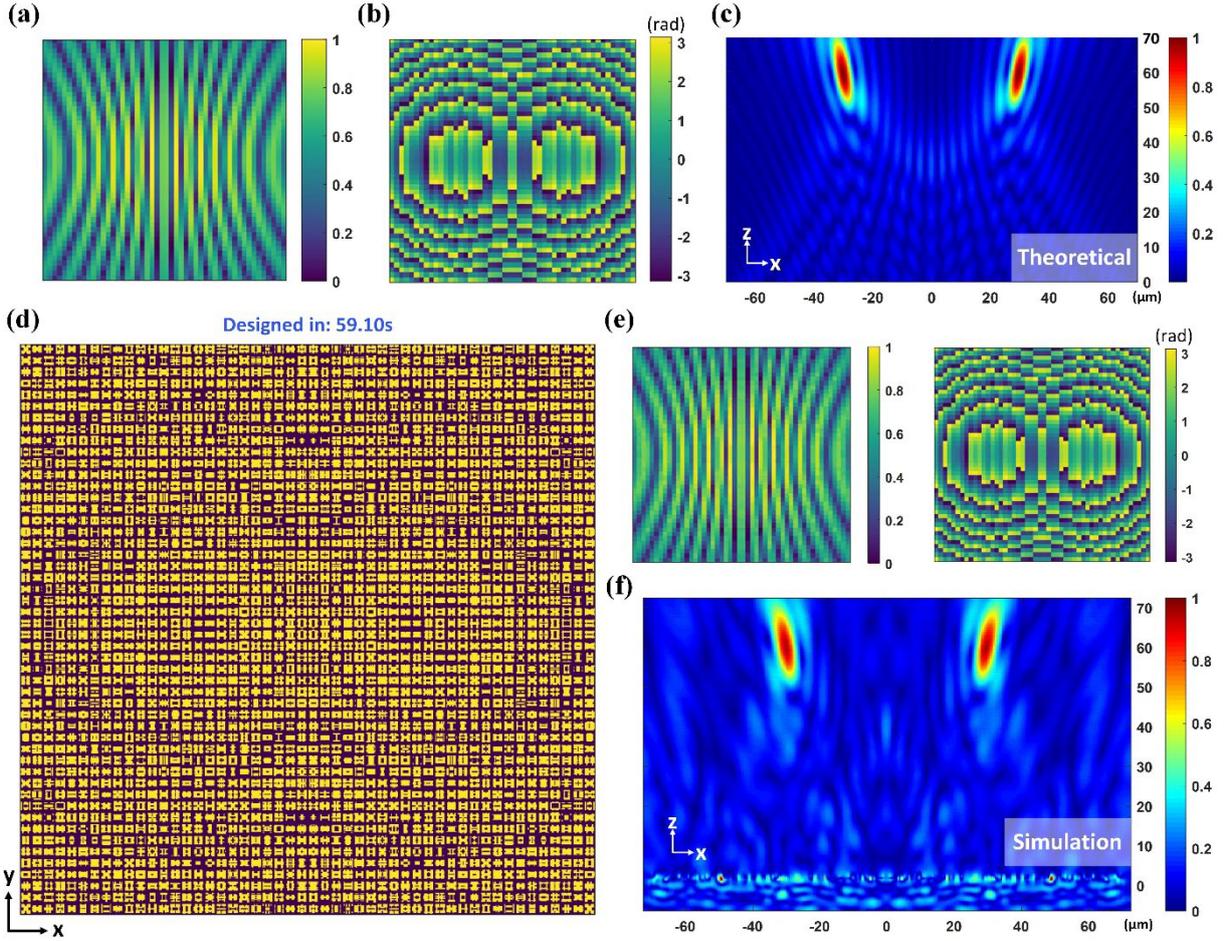

**Figure 3.** A bifocal lens designed with the meta-atom generative network. (a) Target amplitude mask and (b) Target phase mask of the designed lateral bifocal metalens at 50 THz with $f_1 = f_2$ = 60 μm. (c) Theoretical $E_x$ field distributions in the X-Z plane calculated with MATLAB. (d) Metasurface pattern designed by the meta-atom generative network. Since the target E-field is symmetrical along x and y axes, only one quadrant of the metasurface lens was designed, with the rest being generated from symmetry by mirroring along the x and y axes. (e) Amplitude and phase masks for corresponding meta-atoms in (d). (f) Full-wave simulation results of the $E_x$ field in the X-Z plane.

**Multifunctional meta-atom design networks.** Due to the network condition vector's high flexibility, the network structure can be easily adapted to generate meta-atom geometries for implementing multifunctional metasurfaces. For example, by enlarging and rearranging the condition vector: $x = [T_{real}(y_{p1}),\ T_{imag}(y_{p1}), T_{real}(y_{p2}),\ T_{imag}(y_{p2})]$, the network can be configured as a dual-polarization meta-atom design network. Moreover, after being fully trained on the data of complex transmission coefficients associated with two orthogonal linear polarizations (along *x* and *y* axes) (e.g. at 55 THz), the proposed generative network is able to compose meta-atom designs based on four distinct inputs: amplitude and phase responses for incident waves with two orthogonal polarization directions (in this case x-polarized and y-polarized). Notably, with a condition vector containing four different design targets, it's nearly impossible to achieve a high-performance design by using the traditional empirical trial-and-error design approaches, since a slight change in the shape of the meta-atom will affect the four design outcomes simultaneously. With the proposed deep learning approach, we were able to find qualified designs in seconds. Similarly, as for the meta-atom design network, a comprehensive set of phase and amplitude targets for two orthogonal polarizations were chosen to test the performance of the trained network model.

As an example, we set the design goal for a horizontal polarization to a specific amplitude-phase value (0.9, 0°) and gradually varied the targeted phase for the orthogonal polarization from 45° to 315° (with a 90° step). For each amplitude-phase value, we employed the combined network to consecutively generate 100 qualified meta-atom geometries to verify its generation stability and time efficiency for this type of task. The electromagnetic responses of generated meta-atoms were simulated and labeled with red and blue dots in **Figures 4**a-d. We adopted a minimum threshold of ± 0.1 and ± 10° for amplitude and phase errors. The allowed values are outlined in red and blue lines in each polar plot of Fig. 4a-d. Evidently, the stricter design targets have limited the design DOF and reduced the design space, which led to increased similarities between qualified meta-atom shapes in Figs. 4b and 4c. This also increased the difficulty of finding a qualified design, which manifested in a relatively longer pattern generation time (e.g. 32s in Fig. 4a for generating 100 qualified designs) compared to the single-target meta-atom designs shown in Fig. 2.

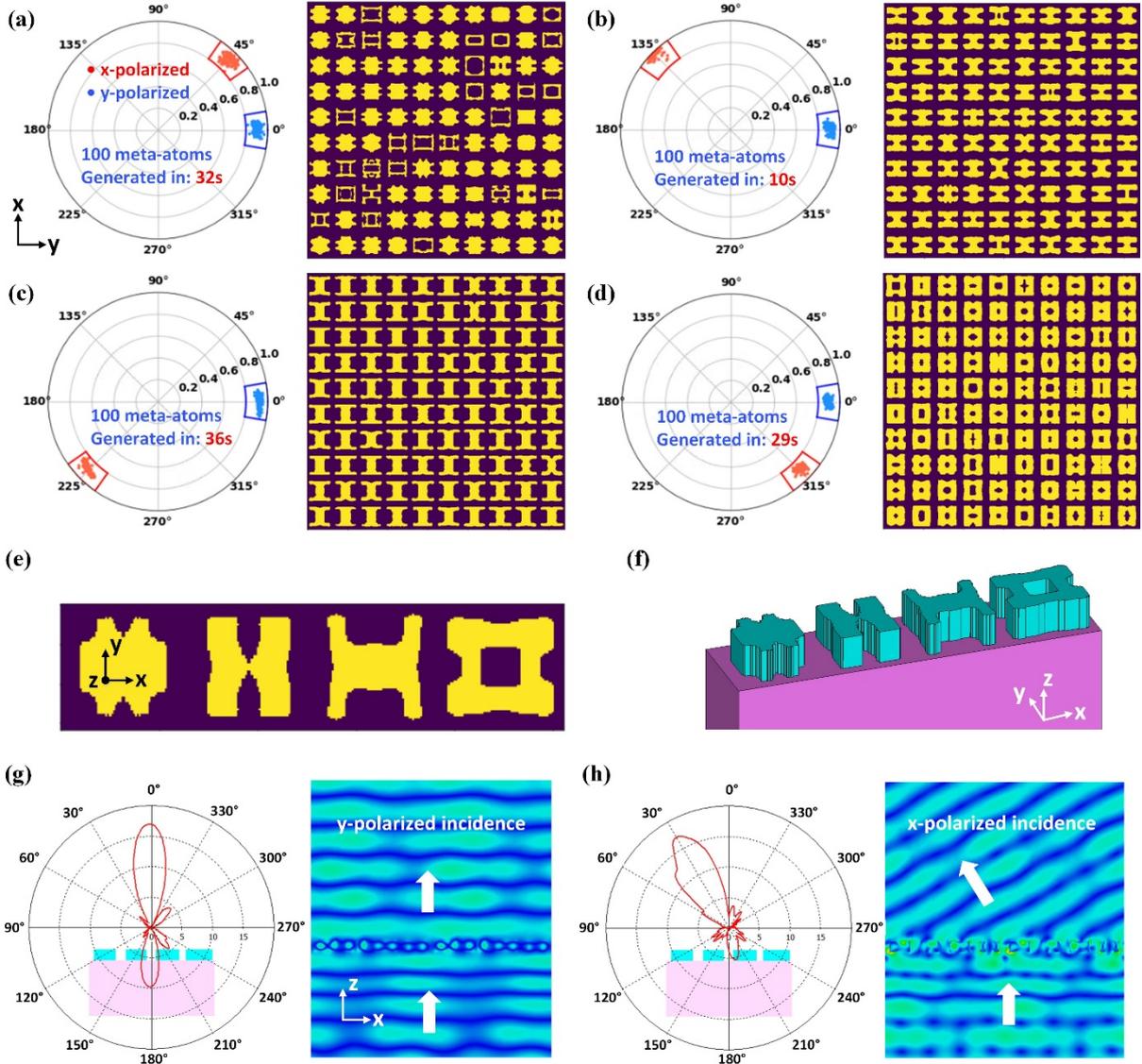

**Figure 4.** Multifunctional meta-atom designs generated using a fully-trained GAN model. EM responses of the generated 100 meta-atoms for two orthogonal polarizations: (a) y-polarization (blue dots): 0.9 + 0°, x-polarization (red dots): 0.9 + 45°; (b) y-polarization: 0.9 + 0°, x-polarization: 0.9 + 135°, (c) y-polarization: 0.9 + 0°, x-polarization: 0.9 + 225°, and (d) y-polarization: 0.9 + 0°, x-polarization: 0.9 + 315°. Red and blue outlines in each polar plot indicate qualified responses under different polarizations, with corresponding patterns highlighted in yellow. (e) Top view and (f) 3D view of a polarization-multiplexed beam deflector assembled with the designed meta-atoms. (g) Simulated E-field angular radiation pattern and Ey field results for a y-polarized plane wave incidence. (h) Simulated E-field angular radiation pattern and Ex field results for an x-polarized plane wave incidence.

After the proposed multifunctional meta-atom design network has been fully trained, we demonstrated its performance through generating several multifunctional meta-device designs working in the mid-IR range. Fig. 4e-f displays the top-view and 3D-view of a polarization-multiplexed beam deflector. The designed beam deflector consists of four meta-atom designs selected from each set of 100 geometries in Fig. 4a-d. The four meta-atoms form a supercell and are tiled along both the x and y directions with periods of 11.2 μm and 2.8 μm, respectively. When illuminated with y-polarized plane waves (Fig. 4g), all meta-atoms have the same phase delays and hence no diffraction beyond the 0th order (specular transmission) appears. With x-polarized plane wave incidence (Fig. 4h), the whole structure acts as a diffractive grating along the x-axis due to the 90-degree-step phase gradient (resembling a traditional blazed grating) to selectively enhance the first diffraction order while suppressing all others. Simulated E-field radiation patterns and E-field amplitude profiles under incidences with different polarization directions are plotted in Fig. 4g and 4h. It is clearly shown that under x-polarized incidence light, most of the optical power is concentrated in the first transmissive diffraction order (at the theoretical deflection angle of 29.14 degrees).

Lastly, a polarization-multiplexed bifocal metalens was designed using the proposed multifunctional meta-atom design network (**Figure 5**). The metalens was designed with a focal length of 60 μm under y-polarized plane wave incidence (Fig. 5a) and 80 μm focal length when illuminated with an x-polarized plane wave (Fig. 5b). The target phase maps of the lens under both x and y-polarized incidence were calculated separately and used as phase inputs for the generative network (first row in Fig. 5d), while the amplitude profiles are kept uniform to maximize the focusing efficiency. Actual phase responses under x- and y-polarized incidences for each meta-atom are simulated and also presented in Fig. 5d (second row) for comparison. The electric fields in the Y-Z plane corresponding to the two different polarization directions were computed by full wave simulation and are plotted in Fig. 5e, where a clear focal spot at the desired focal length is observed for each case. The electric fields along the optical axis were also simulated and plotted in Fig. 5f, where two distinct electric field peaks can be clearly observed at 60 μm and 80 μm, respectively. Similarly, we can also easily design a polarization-independent lens utilizing the same generative network (Supporting Information Section 5).

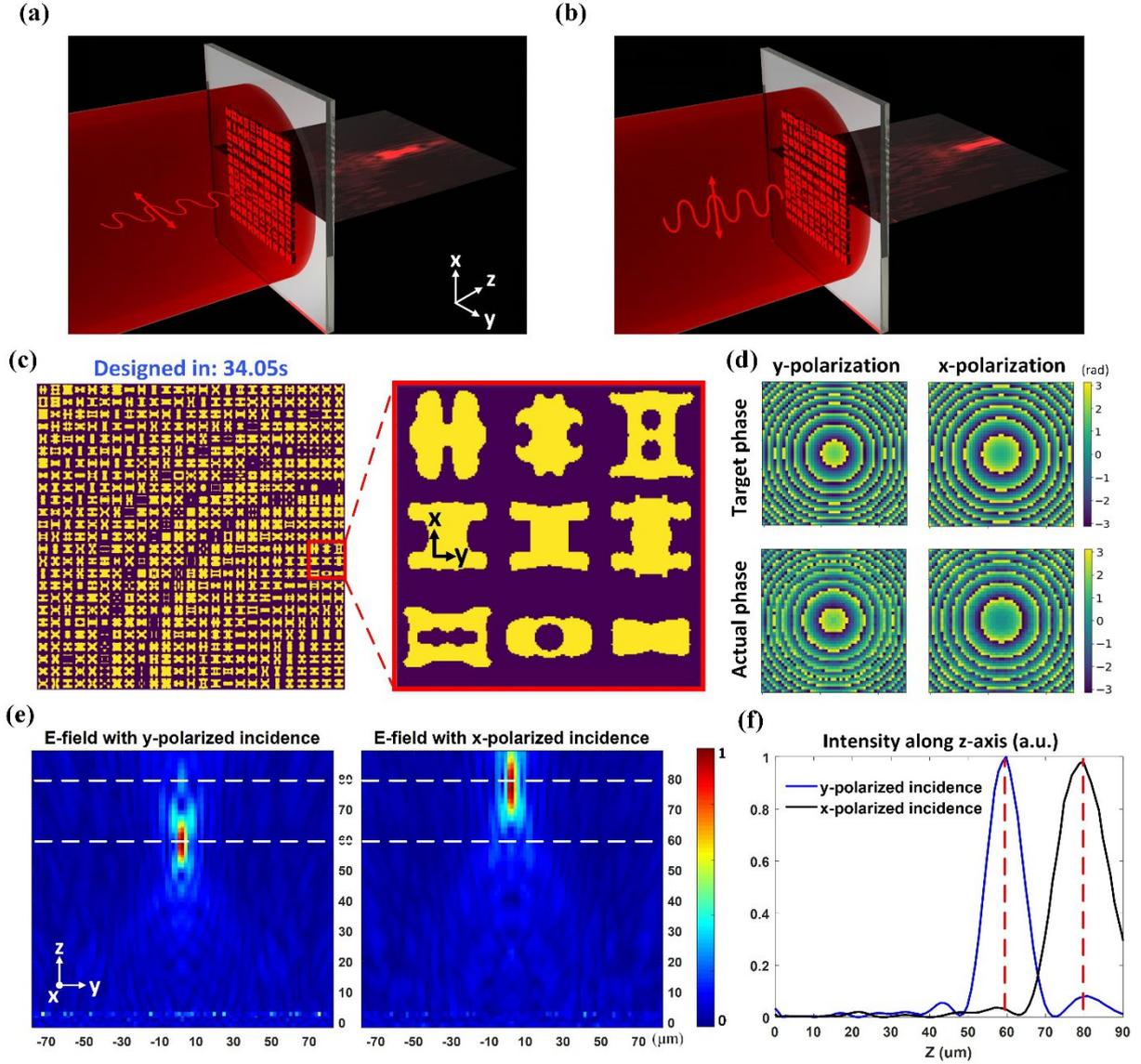

**Figure 5.** A bifocal metasurface lens designed with the dual-polarization meta-atom generative network. The lens has a 140 μm × 140 μm aperture size, containing 50 × 50 meta-atoms in total. 3D illustration demonstrating the metalens functionality under (a) y- and (b) x-polarized incident light (c) Metasurface pattern designed with the dual-polarization meta-atom generative network. Only the top left quadrant of the lens was generated, with the rest being duplicated according to the design symmetry. (d) Target (first row) and actual simulated (second row) phase masks for the bifocal lens for each polarization incidence. (e) Full-wave simulated E-field in y-z plane under y and x-polarized incidence, respectively. The focal spot is shifted from z = 60 μm to z = 80 μm when the polarization direction switched from y to x. (f) Simulated E-field intensity along the optical axis.

**Discussion and conclusion.** The ability of the proposed network to precisely achieve the multifunctional design goals, as shown in Fig. 2 and 4, along with the results of the assembled metasurface devices shown in Fig. 3-5, have verified the utility, power, and ease of use for this approach. We believe that our approach provides a much more effective means of meta-device design compared to trial-and-error or global optimization design approaches since: 1) once the training is completed, our GAN generates designs with stable performances with almost zero time cost and does not depend on the initial guess of the meta-atoms' shapes and parameters. Furthermore, meta-atom designs derived with the proposed approach can be specified as high-quality initial designs and further refined with optimization algorithms, which provides a potential solution for addressing the local minima optimization problems. 2) By training the proposed GAN with data collected from meta-atoms with various refractive indexes, thicknesses, lattice sizes, 2D cross sections and loss factors, the proposed approach can be easily extended to a variety of metasurface platforms based on different materials and fabrication processes. 3) The condition vector in the proposed GANs are highly versatile and can be easily customized into various multifunctional design goals, which remains a major challenge to traditional design approaches. In addition to the polarization-multiplexed metasurface designs presented in the paper, the proposed network structure can also be easily customized for designing other multifunctional devices including multi-band meta-atom/metasurfaces, wideband meta-filters, tunable meta-devices and many more.

In addition to its use for meta-device design, the proposed GAN may be a useful tool for topological analysis of meta-atom structures. Examination of classes of structures generated by the proposed network that share a particular EM response (as shown in Fig. 2) [51, 52] can lead to the discovery of underlying physical characteristics. By processing the image through several convolutional layers, the neural network can uncover the common traits of these designs; these common traits can be used to categorize the designs into the same conditional distribution, which is highly non-intuitive. Designs with inclined edges and round corners are generated (Fig.

2a-f), despite that training datasets collected with the "Needle Drop" approach are all composed of rectangles with straight sharp edges and limited to only 28 × 28 resolution. This result highlights the exploratory (learning and composing) nature of the proposed network, which utilizes the increased 64 × 64 image resolution to yield design details that are not included in the training data and transcend training data limitations.

To conclude, we have proposed a metasurface design network based on the conditional WGAN architecture that is capable of efficiently producing numerous multifunctional meta-device designs on demand. The fully-trained network demonstrated this capability through several example designs, including a bifocal metalens, a polarization-multiplexed beam deflector, a polarization-multiplexed metalens and a polarization-independent metalens. Excellent agreement has been achieved between design targets and generated device performance for each design. Furthermore, we suggest that the proposed network can be used as a tool for topological analysis in uncovering shared physical features within groups of similar electromagnetic responses. We envision that this deep-learning-based design approach can be readily applied beyond multifunctional metasurfaces/meta-atoms to various types of other multifunctional electromagnetic devices, such as microwave components, antennas, and integrated optical circuits.

**Acknowledgements**
This work was funded under Defense Advanced Research Projects Agency Defense Sciences Office (DSO) Program: EXTREME Optics and Imaging (EXTREME) under Agreement No. HR00111720029.

# Supporting Information

## Title: Multifunctional Metasurface Design with a Generative Adversarial Network

*Sensong An[1], Bowen Zheng[1], Hong Tang[1], Mikhail Y. Shalaginov[2], Li Zhou[1], Hang Li[1], Myungkoo Kang[3], Kathleen A. Richardson[3], Tian Gu[2], Juejun Hu[2], Clayton Fowler[1,*], Hualiang Zhang[1,*]*

## 1. Detailed network architecture

Fig. S1 illustrates the network architecture of the proposed neural network, with the data flow details included. Input conditions, such as frequency-dependent amplitude and phase responses, polarization-dependence, and/or material states (e.g. for tunable materials such as phase-change materials) were combined with randomly generated noise. The combined input was fed into the generator to produce fake meta-atom samples that resemble real ones well enough to confuse the discriminator. The discriminator learns by maximizing the Wasserstein distance between the fake and real samples. Both modules approach the real data distribution through parameter tuning during this adversarial process. More detailed architectures of the generator and discriminator were shown in Fig. S1b and S1c, respectively.

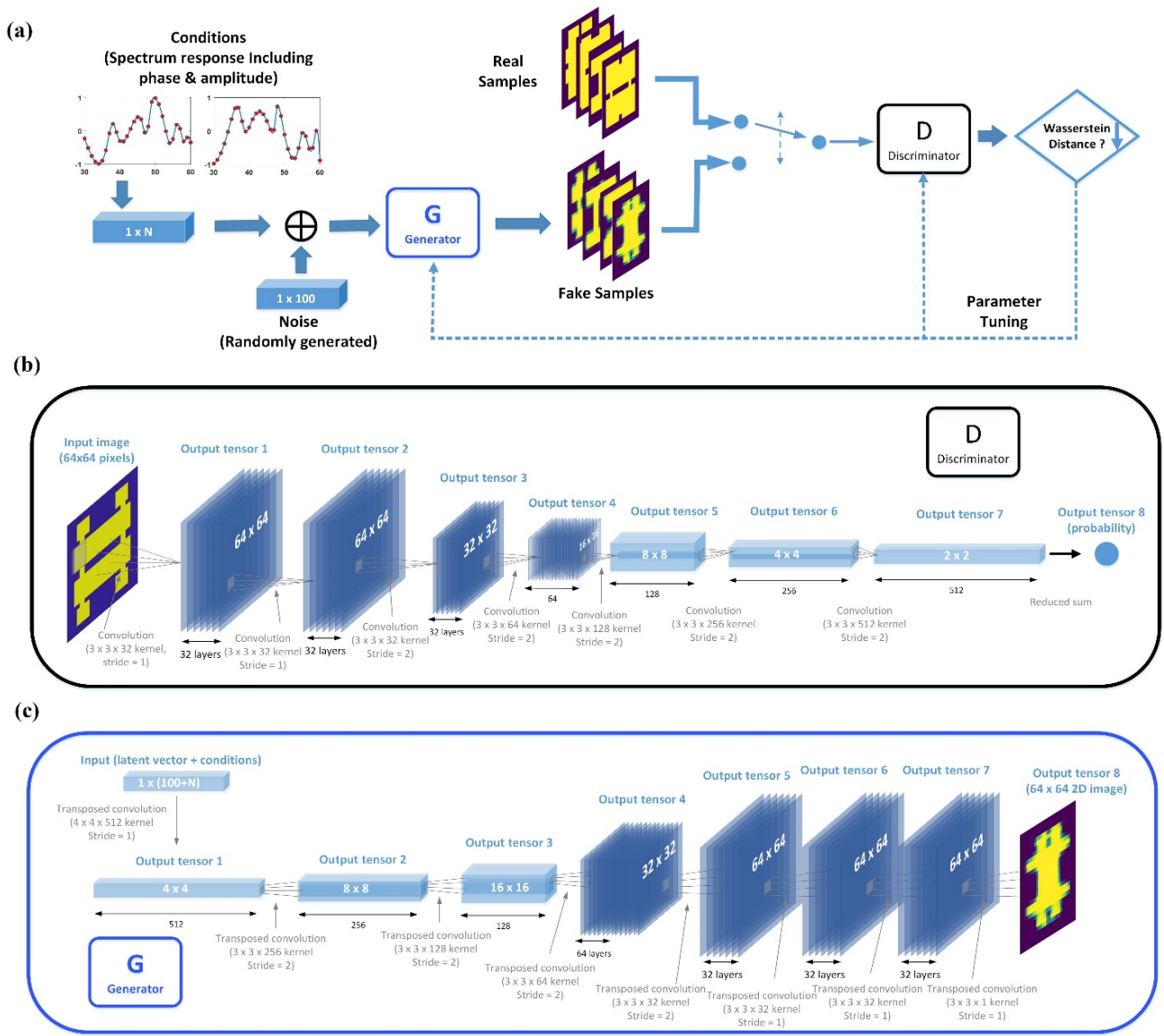

**Figure S1. Detailed network architectures of the GAN.** (a) Illustration of the training process of the GAN. (b) Detailed network structure of the discriminator. The discriminator consists of seven consecutive convolution layers. The output of each layer is batch-normalized and passed through a ReLU activation function before being passed on to the next layer. The initially planar dimensions of an input are decreased while the depths are increased via (2, 2) stride convolutional layers. The output tensor of layer #7 is flattened into a 1D array and the reduced sum is calculated to represent its Wasserstein distance. (c) Detailed network structure of the generator. Conversely, the generator consists of eight consecutive transposed convolution layers for which the depth of an output tensor is decreased while gradually being flattened into a 2D meta-atom image. The output of each layer is batch-normalized and passed through a Leaky ReLU activation function. After the last transposed convolutional layer, a *tanh* activation function generates a 2D image representing the meta-atom design. Details of each output tensor, shapes of the (transposed) convolutional kernels and strides used during convolutions are given in the figure.

For more design DOF, despite that the original input images (with 28 × 28 pixels resolution) that are used to sketch the 2D shape of the meta-atoms from the training dataset, the input images were all rescaled into 64 × 64 pixel images before they were processed using convolutional layers. The generator and the discriminator were trained alternately: after the discriminator was trained with 3 different batches of training data, values of the hidden neurons in each convolutional layer are updated and fixed. The generator was then trained with a new batch of training data with the help of the most up-to-date discriminator, and so on, until training completes. Details, including dimensions of each output tensor, shapes of the convolutional kernels and strides used during convolutions are given in Fig. S1b and S1c.

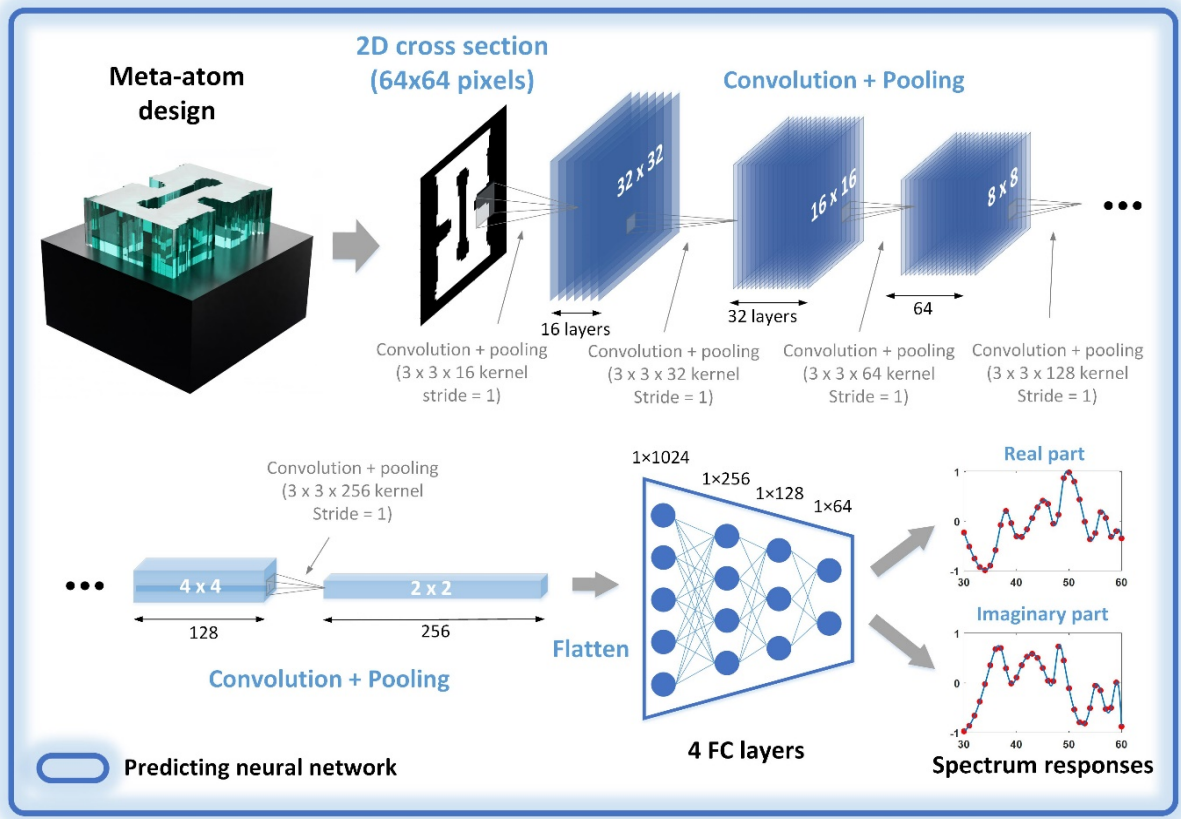

**Figure S2.   Detailed network architectures of the predicting neural network (PNN).** Details including dimensions of each output tensors, shapes of the (transposed) convolutional kernels and strides used during convolutions are given in the figure.

Following the approaches introduced in [49], we constructed two PNNs (Fig. S2) to predict the real and imaginary parts of the transmission spectrum, respectively. Transmissive amplitude and phase are then derived using the predicted real and imaginary parts. Detailed network architecture of the PNN cascaded to the GAN is shown in Fig. S2. The PNN was constructed based on a convolutional neural network (CNN) architecture. It functions as a critic and examines the performance of the designs generated by the proposed GAN. Specifically, the PNN is able to precisely predict the transmission spectrum of free-form meta-atom designs within the frequency range of 30 to 60 THz. In contrast to full-wave simulation tools, the PNN characterizes the meta-atoms on a one-time calculation basis and, thus, significantly speeds up the whole design process.

## 2. Training data collection

Without loss of generality, the all-dielectric meta-atom consists of a 1 μm thick dielectric component (preferably with a high refractive index, $n_1$. In this case $n_1 = 5$) sitting on a dielectric substrate (preferably with a low refractive index, $n_2$. In this case $n_2 = 1.4$) with a unit cell size of $2.8 \times 2.8$ μm$^2$ (Fig. S3a). The 2D pattern of each meta-atom was generated with the "needle drop" approach using the numerical computing tool MATLAB. Several (3 to 7) rectangular bars, with a minimum generative resolution of 0.1 μm, were randomly generated and placed together within a square lattice to form random patterns (Fig. S3b). To minimize inter-cell coupling, a minimum spacing of 0.4 μm was applied between adjacent meta-atoms. To speed up the data-collection process, the all-dielectric components are only generated in the top left quadrant of each unit cell and then symmetrically replicated along $x$ and $y$ axes to form the whole pattern. A set of meta-atoms generated in this manner is guaranteed to possess polarization-diverse performance.

The full-wave electromagnetic simulations were performed using a commercial FEM simulation tool CST. For each meta-atom, perfect electric conducting surface ($E_t = 0$) and perfect magnetic conducting surface ($H_t = 0$) boundary conditions were employed to calculate the transmission and phase shift of a square lattice structure. Open boundaries are applied along both the negative and positive $z$ directions, while an $x$-polarized plane wave was illuminated from the substrate side for each meta-atom. To further accelerate the full-wave simulations, $E_t = 0$ and $H_t = 0$ symmetry planes were applied in the center $y$-$z$ plane and $x$-$z$ plane for each meta-atom, respectively. A total number of 69,000 meta-atoms with different shapes were generated and simulated to find their wide-spectrum phase and amplitude responses. These simulations were performed on eight servers running in parallel. The data collection process was completed in 3 days. After removing similar patterns (to speed up the training), 29,000 meta-atom structures were selected and documented for further training.

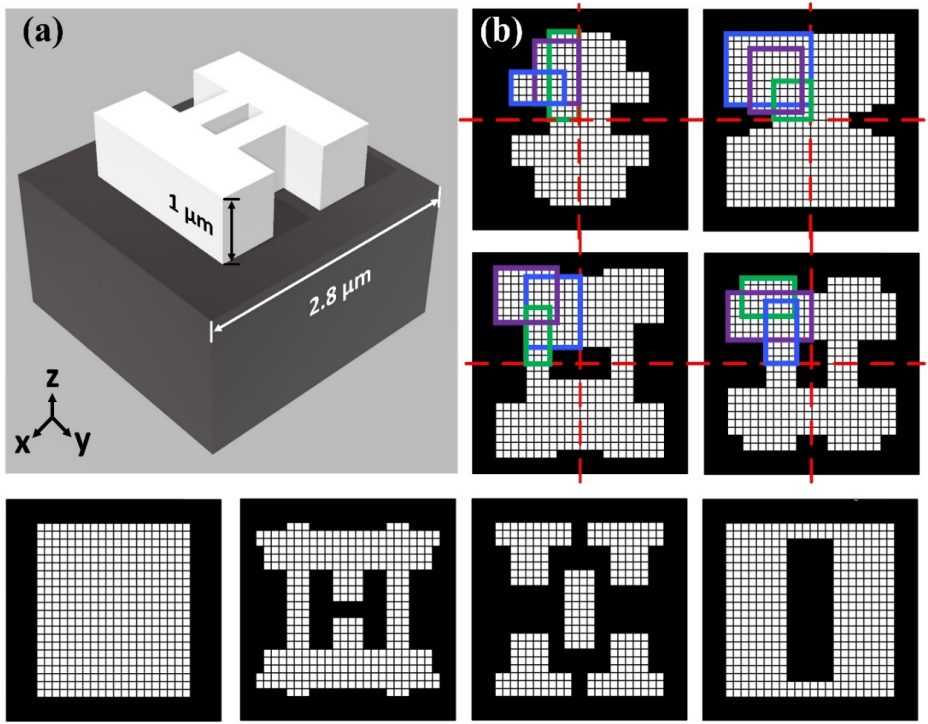

**Figure S3. Training data collection process.** (a) 3D view of a generated meta-atom of arbitrary shape. Lattice constant for each meta-atom cell is 2.8 μm, meta-atom height is fixed to be 1 μm. White and black colors represent high- and low-index (substrate) dielectric components. **(b)** Demonstration of the pattern generation process. 2D patterns in x-y plane are meshed, each mesh pixel has a dimension of 0.1 by 0.1 μm². Rectangles outlined in different colors represent distinct high-index "needles" that were randomly generated and dropped on the top-left quadrant of the substrate canvas. Patterns were completed by mirroring the pattern along the *x* and *y* axes.

## 3. Customized gradient-penalty method

The Wasserstein distance is only accurate when the discriminator is a 1-Lipschitz function.[45] To enforce this constraint, the original WGAN applied a simple, but rough, value clipping to restrict the maximum weight value in each layer of the discriminator. Instead, WGAN-GP uses a gradient penalty term to ensure that the norm of its gradients is equal to 1 almost everywhere [46] so that the discriminator is 1-Lipschitz. Conditional-Wasserstein distance with a gradient penalty term can be represented as:

$$W(P_{data}, P_G) \approx \sup_{\|D\|_L \leq 1} \Big\{ \mathbb{E}_{y \sim P_{data}}[D(y|x)] - \mathbb{E}_{y' \sim P_G}[D(y'|x)]$$

$$- \lambda \mathbb{E}_{y \sim P_{penalty}}[\max(0, \|\nabla_y(D(y))\| - 1)] \Big\} \tag{S1}$$

Traditional WGAN-GP randomly interpolates between network generated patterns $P_{generator}$ and real sample patterns $P_{data}$ to generate gradient penalty samples $P_{penalty}$, as shown in Fig. S4a. The interpolation method has an important advantage: as the training progresses and $P_{generator}$ approaches $P_{data}$, the gradient norm of this more widespread distribution $P_{penalty}$, instead of the real sample distribution $P_{data}$, satisfies the Lipschitz constraint, and we can thus conclude that the discriminator is 1-Lipschitz almost everywhere within $P_{data}$.

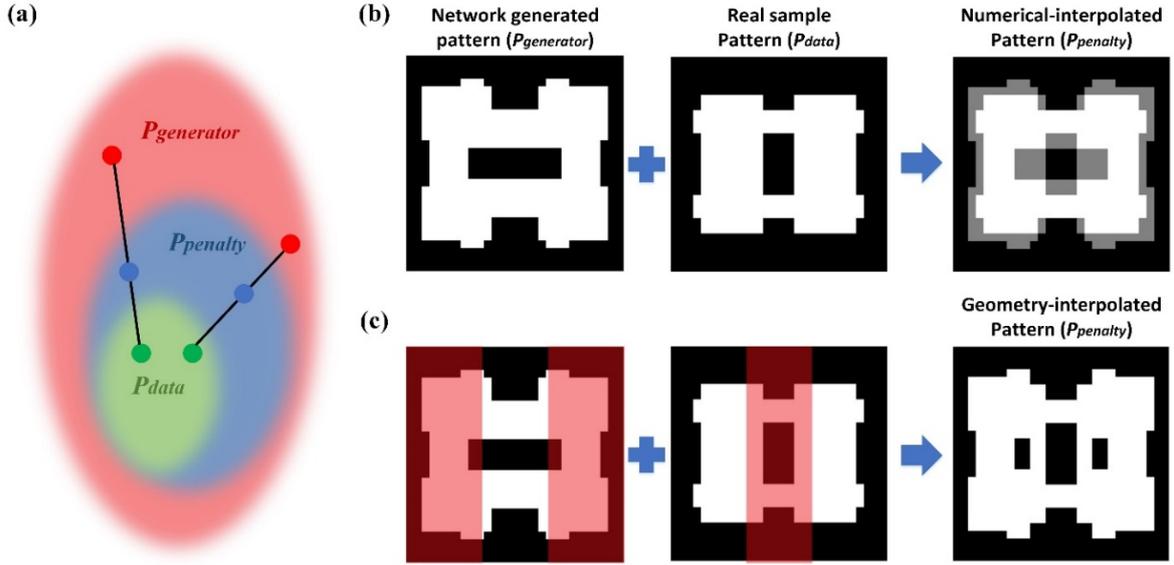

**Figure S4. A novel interpolation method for customized gradient penalty.** (a) Schematic diagram of the random interpolation process in WGAN-GP. The network randomly interpolates between $P_{generator}$ and $P_{data}$ to get $P_{penalty}$. (b) Numerical interpolation methods employed by traditional WGAN-GPs. The meta-atom patterns are binarized, such that white represents "1" and black represents "0". (c) The proposed novel geometry interpolation method. Random proportions (marked in red) were taken from both fake samples and real samples and later combined into a new pattern.

In our case, with meta-atom patterns as target design goals, the generated outputs can be converted into binary images consisting of 1's that represent the dielectric material and 0's that represent voids. The conventional numerical interpolation process is not applicable for the meta-atom discriminator, because generated values between "0" and "1" don't correspond to any physical structures (Fig. S4b). As a result, the discriminator that is trained to satisfy the Lipschitz constraint for this $P_{penalty}$ is intuitively challenged to yield stable Wasserstein distance results for real samples from $P_{data}$ during the training. We therefore employed a novel geometry interpolation method that combines random geometry portions from both $P_{generator}$ and $P_{data}$ to form the sample in $P_{penalty}$ (Fig. S4c). The interpolated results obtained in this manner fully characterize the samples between $P_{generator}$ and $P_{data}$. This unique interpolation method also allows the generator to extrapolate and explore the ground truth distribution when the training data is insufficient to cover the whole design space. Training experiments which validate the training stability, design accuracy and extrapolation capability of the proposed gradient-penalty method are presented in Section 4.

### 4. Hyperparameters and training curves

**Table S1. Hyperparameters used in the training of GANs and PNNs.**

| Hyperparameters | Meta-atom design network | Dual-polarization meta-atom design network | PNN for real part | PNN for imaginary part |
|---|---|---|---|---|
| **Training set size** | *29,000* | *29,000* | *69,000* | *69,000* |
| **Optimizer (learning rate)** | Adam *(1e-4)* | Adam *(1e-4)* | Adam *(1e-4)* | Adam *(1e-4)* |
| **Batch size** | *64* | *64* | *256* | *256* |
| **Batch Norm.** | Yes | Yes | No | No |
| **Nonlinear activations** | ReLU for the discriminator, Leaky ReLU (alpha = 0.2) for the generator, *tanh* for the generator's last layer | | ReLU | |
| **Penalty coefficient** | *10* | *10* | NA | NA |
| **Iterations (time)** | *3,000 (72* h) | *3,000 (72* h) | *10,000 (6* h) | *10,000 (6* h) |

Hyperparameters used during training are shown in Table S1. Training curves for the meta-atom design network, dual-polarization meta-atom design network, and PNN are shown in Fig. S5. Despite the slightly different structures and training data fed to these networks, their generator losses all decrease gradually while discriminator losses remain constant. As shown in Fig. S5 (a-b), after approximately 3,000 epochs of training, each network converged to a point that both generator and discriminator loss stabilized, which means that the generator is able to generate samples that are close enough to the real samples that the discriminator is unable to differentiate between real and fake.

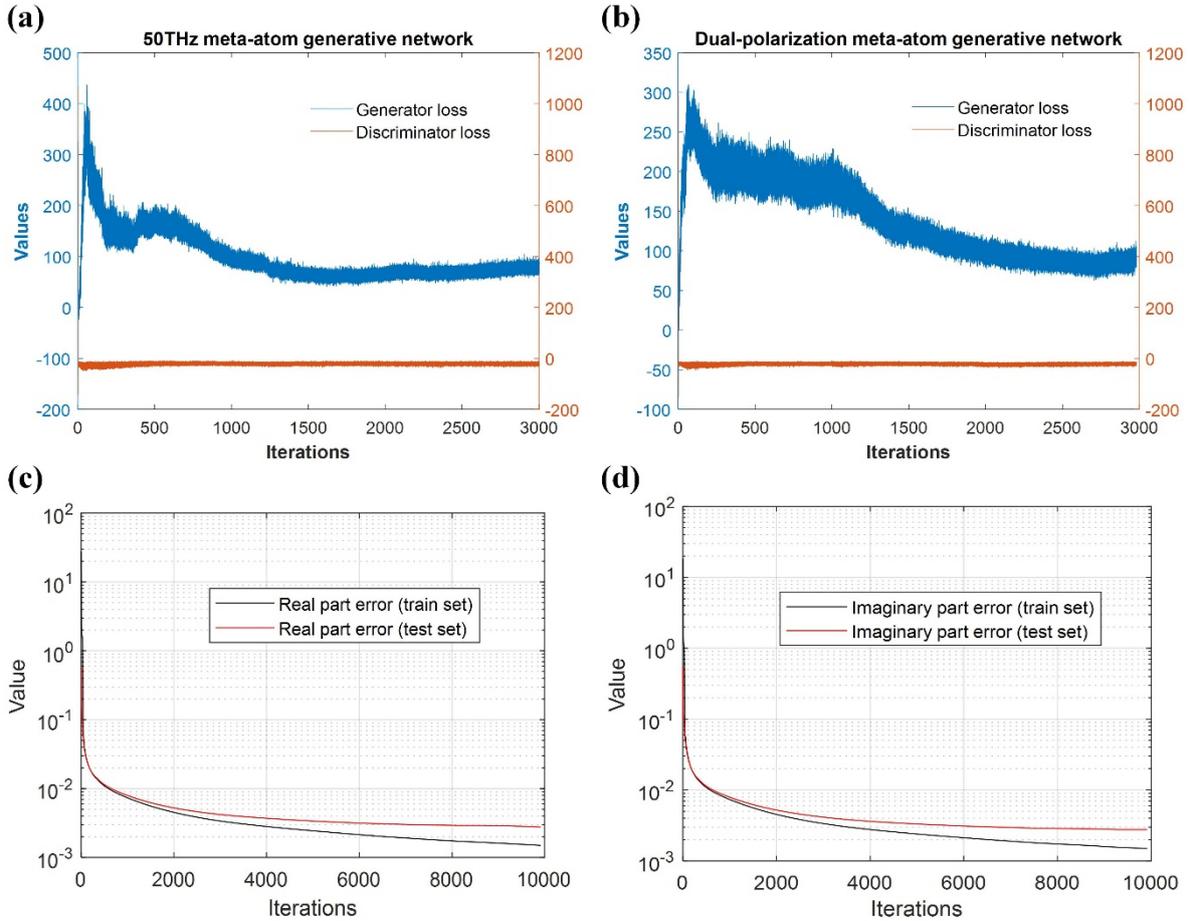

**Figure S5. Training losses of the two GANs and the PNN proposed in the paper.** (a) The meta-atom generative network operating at 50 THz. (b) The dual-polarization meta-atom generative network working at 55 THz. (c) PNN real part prediction. (d) PNN imaginary part prediction.

Training curves for the two constructed PNNs were included in Fig. S5(c-d). Both networks are trained with the same 69,000 groups of training data collected for the training of the GANs. Both networks converged well after 10,000 iterations.

To better visualize how the network learned the meta-atom design principles and actually "evolved" during the training process, we recorded the network models during the training process and employed several half-trained models to design the same bifocal metalens, presented in Fig. 3, and tested their performance by numerical simulations. Four different sets of bifocal metalens designs based on GAN models derived after 1, 2, 100 and 3,000 training iterations are presented. The designed metalenses, along with their full-wave simulated E-fields, are plotted in Fig. S6. Interestingly, at the beginning of the training process, the generated meta-atoms designs have similar shapes with large volumes and unclear boundaries (Fig. S6a). As the training proceeds, the GAN models start to generate meta-atom patterns with more diverse shapes and refined details (Fig. S6b-d). The corresponding E-field distributions also gradually converged to two sharp focal spots (Fig. S6d), as desired, which firmly confirm the increasing learning capability of the proposed GAN model during the training process.

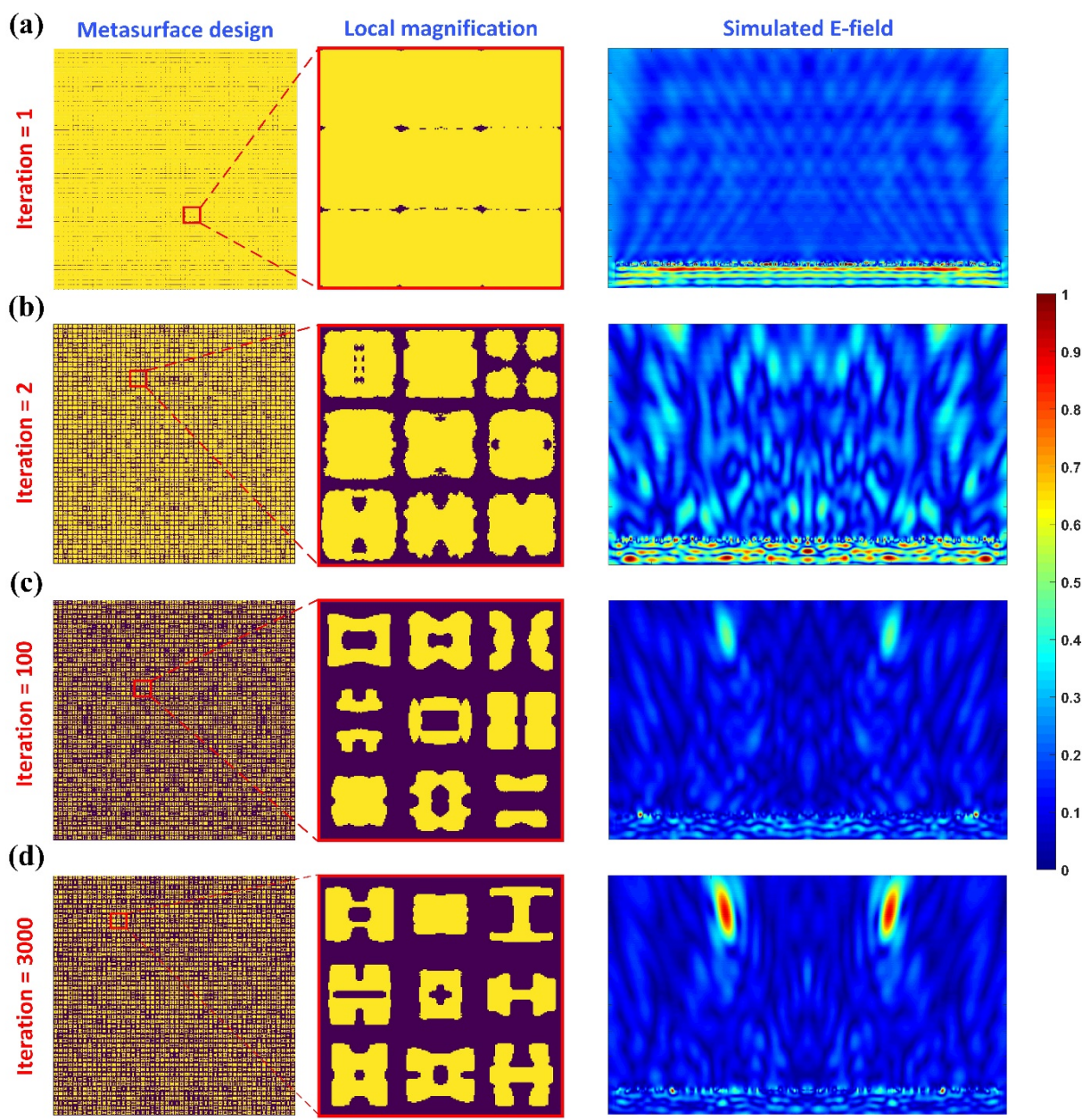

**Figure S6. Visualization of the GAN training process.** Bifocal metalenses with target amplitude and phase maps shown in Fig. 3a and 3b, designed using GAN models trained for (a) 1 iteration, (b) 2 iterations, (c) 100 iterations and (d) 3,000 iterations. Several meta-atoms from each metalens design are magnified for a clear view. Numerically simulated E-field of each metalens were performed using CST and plotted on the right side of each subplot.

## 5. Polarization-independent metalens design

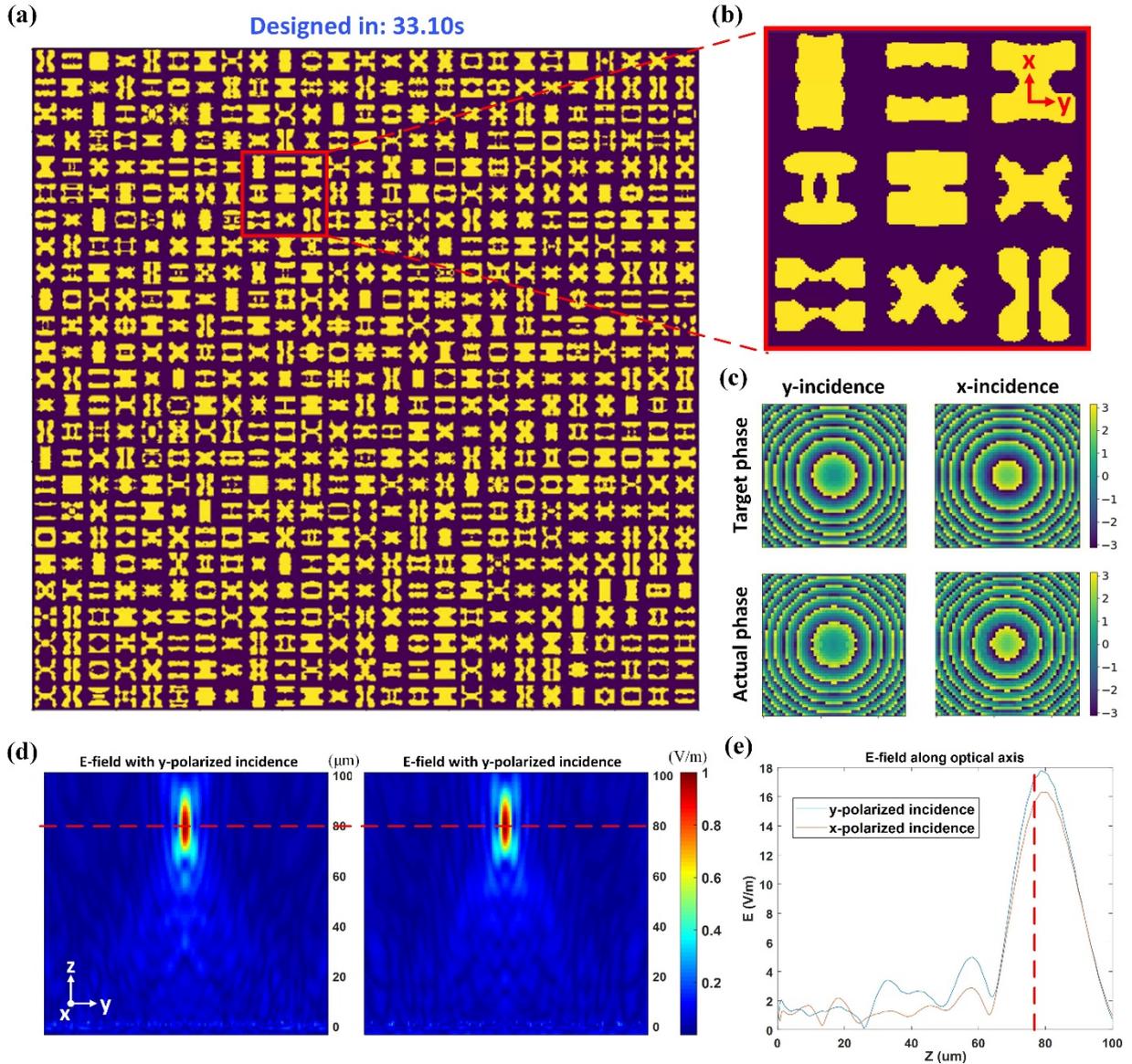

**Figure S7. A polarization-independent focusing lens designed with the dual-polarization meta-atom generative network.** (a) Metasurface pattern designed with the dual-polarization meta-atom generative network. Part of the metasurface device circled in red lines was enlarged in (b) for a clear view. (c) Element-wise phase and amplitude responses of each designed meta-atom under y and x polarized incidences. (d) Full wave simulated amplitude of E-field in y-z plane under two orthogonal polarization incidences. Focal length remained 80 μm while polarization direction was switched. (e) Full-wave simulated E-field along optical axis under two orthogonal polarization directions.

To further explore the versatility of the multifunctional meta-atom design network, we utilized our dual-polarization meta-atom generative network to design a polarization-insensitive transmissive focal lens with an equal focal length of 80 μm for both polarizations. One way to achieve this goal is enforcing that the phase shifts for *x* and *y* polarization are identical, which primarily results in structures that feature 4-fold rotational symmetry, in accordance with their polarization-insensitive nature. The 4-fold rotational symmetry requirement can be relaxed by considering that the relative phase difference between two polarization states need not be zero, but simply maintained constant. Following this approach, with the help of dual-polarization meta-atom generative network, a 90 degree constant phase bias (difference) was added to the lens' phase mask under x-polarized plane wave incidence (versus that under y-polarized incidence). For each single cell in the metasurface lens, its target phase profiles under both polarization were designated as input of the network and one qualified design was generated to assemble the metalens (Fig. S7a). The full wave simulated electric fields in y-z plane for the whole lens in Fig. S7a are plotted in Fig. S7d. The E-fields along the optical axis in both cases share the same focal length of 80 μm with near-equal magnitude (Fig. S7e), validating the efficacy of proposed design approach.